\documentclass[journal=jacsat,manuscript=article]{achemso}

\usepackage{graphics,latexsym}
\usepackage{color}
\usepackage{setspace} 
\usepackage{amsmath}
\usepackage{amssymb}
\usepackage{epstopdf}

\title{Quantum State Tuning of Energy Transfer in a Correlated Environment}
\author{Francesca Fassioli}
\affiliation{Department of Physics, University of Oxford, Clarendon Laboratory, Parks Road, Oxford OX1 3PU, UK}
\author{Ahsan Nazir}
\affiliation{Department of Physics and Astronomy, University College London, Gower Street, London WC1E 6BT, UK}
\author{Alexandra Olaya-Castro}\email{a.olaya@ucl.ac.uk}
\affiliation{Department of Physics and Astronomy, University College London, Gower Street, London WC1E 6BT, UK}

\begin{document}

\begin{abstract} 
We investigate multichromophoric energy transfer allowing for bath-induced fluctuations at different 
sites to be correlated. As a prototype system we consider a light-harvesting antenna surrounding  a reaction center. 
We show that the interplay between quantum coherence and correlated fluctuations can generate a room 
temperature transfer process featuring a marked dependence on the degree of symmetry and delocalization of the initial 
exciton state.  Our work illustrates how these quantum features could support fine tuning of energy transfer efficiencies in 
closely-packed natural and artificial light-harvesting complexes.
\end{abstract}


\maketitle
\newpage

Remarkable experimental advances in multichromophoric assemblies have raised the fascinating possibility that quantum 
coherent dynamics may be relevant in photosynthetic energy transfer,  
even at room temperature \cite{engel07, lee07, collini09, collini10, scholes10}. Key to the survival of quantum coherence in this 
temperature regime seems to be the emergence of correlated energetic fluctuations between different chromophores 
\cite{nazir09}. This could be due, for instance, to closely spaced pigments
sharing the same environmental modes \cite{lee07,collini09}. There is thus a growing need for theoretical descriptions 
of energy transfer beyond the common assumption of independent environments for each pigment \cite{beljonne09, cheng09}. 
Indeed, a number of studies have addressed this issue \cite{renger98,yu08,hennebicq09,rebentrost09,plenio08,caruso09,nazir09},  
though the subtle interplay between coherence and correlated fluctuations in influencing transfer efficiencies remains 
largely unexplored.  

The prospect of environment-protected, room-temperature quantum coherence in photosynthetic systems has revived the 
longstanding question on the functional role such quantum phenomena may play. 
In particular, it is natural to ask whether the interplay between coherence and correlated fluctuations provides the 
system with an {\it efficiency control mechanism} that it could not otherwise achieve \cite{cheng06}. In this work, we 
address this question with the aid of a model of multichromophoric energy transfer under the influence of spatially 
correlated energetic fluctuations. We show that a fundamental signature of  
the interplay between coherence and correlations is a quantum yield with a conspicuous dependence on the phase 
information embedded in the initial exciton state. In particular, the energy transfer efficiency can distinguish the 
degree of symmetry or asymmetry of the initial exciton state, as well as its delocalization length. Such a dependence 
produces an ordering of the excitonic eigenstates from higher to lower efficiency as a function of increasing energy, a 
feature which does not exist under local environments. Hence, coherence and correlations can indeed equip large, densely 
packed photosynthetic systems with a unique mechanism to control their transfer efficiency.

\begin{figure}
\centering
\resizebox{6.0cm}{!}{
\includegraphics*{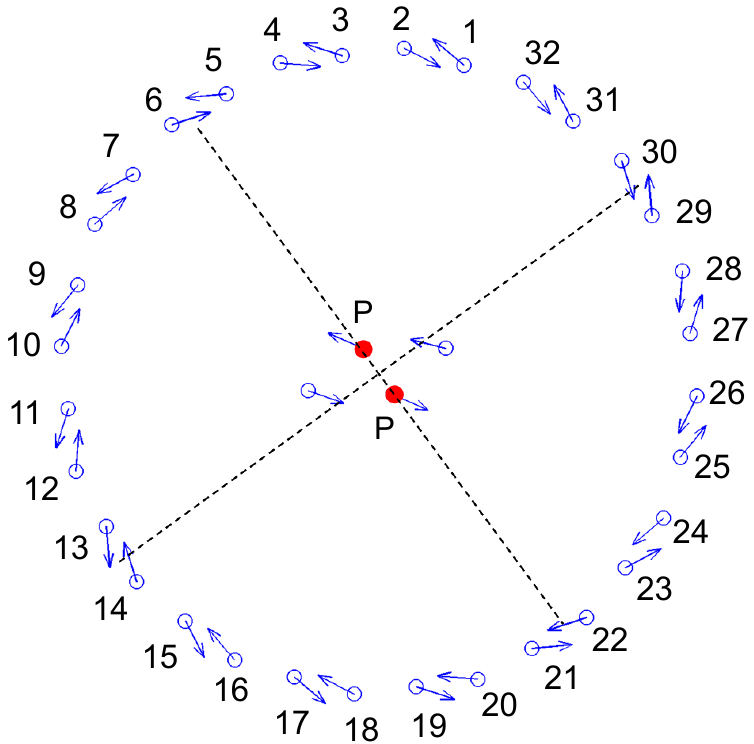}}
\caption{Schematic of the LH1-RC core of purple bacteria based on Ref. \cite{hu97}. Arrows indicate the 
transition dipole moments for each bacteriochorophyll (Bchl), as given in Ref. \cite{hu98}. The special acceptor pair in 
the center (red dots) defines a set of preferred axes of symmetry for the electronic couplings between the center and 
the LH1 Bchls \cite{alexandra08}.}
\label{fig:ring}
\end{figure}

As a prototype system we consider a light-harvesting 1 (LH1) antenna connected to a reaction center (RC) as in purple 
bacteria \cite{hu98, hu97,  damjanovi00, richter07, alexandra08} (see \ref{fig:ring}).  We focus on this unit as it 
exhibits key features, such as symmetry in its geometry and interactions, representative of LH units present in a large 
variety of natural species \cite{simon07} and artificial systems \cite{womick09}; such symmetries are expected to play 
an important role in the energy transfer process.  Evidence of coherence in the LH1-RC core is sparse, though its 
excitonic states can be delocalized over a large number of pigments at low temperatures \cite{ketelaars02}. Moreover, it 
has been shown that correlated energy fluctuations among several Bchls in the LH1 ring must be included to accurately 
reproduce spectroscopic properties \cite{rutkauskas06}, pointing again to the importance of environmental correlations 
in these densely-packed pigment-protein aggregates.   
The Hamiltonian describing the electronic excited states of the $M$ strongly interacting pigments of the LH1-RC core  is 
given by $H_S=~\sum_{m=1}^{M}(E_m +\delta E_m)\sigma^{+}_{m} \sigma^{-}_{m}+\sum_{n>m}V_{mn}(\sigma^{+}_{m} 
\sigma^{-}_{n} + \sigma^{+}_{n} \sigma^{-}_{m})$, with $\hbar=1$. The operator $\sigma^{+}_{m}$ creates a localized 
excitation at site $m$, with an energy  separated into an ensemble average $E_m$, and a deviation $\delta E_m$ that 
accounts for diagonal static disorder (see supporting information).    
On-site energies $E_m$ and electronic couplings $V_{mn}$ have been derived  from Refs. \cite{hu97, hu98}  and 
they reproduce the observed exciton spectra \cite{ExcitonSpectra}.
In contrast to the common assumption of local, independent environmental modes in contact with each pigment, we consider 
here a single thermal bath described by $H_B=\sum_{\bf k}\omega_{\bf k} b_{\bf k}^{\dag}b_{\bf k}$, where $b_{\bf k}^{\dag}$ 
($b_{\bf k}$) are the creation (annihilation) operators of bosonic modes of frequency $\omega_{\bf k}$.
Each pigment is assumed to have a {\it position dependent} coupling to the bath of the form $(g_m)_{\bf k}=|(g_m)_{\bf 
k}|e^{i{\bf{k}}\cdot{\bf{r}}_m}$, with ${\bf r}_m$ the position of site $m$ with respect to each phonon mode 
\cite{doll06}. This position dependence  will give rise to correlated energy  fluctuations. For simplicity, we consider 
isotropic and identical coupling strengths for all sites, $|(g_m)_{\bf k}|\equiv|g_{\bf k}|$, with associated 
single-site spectral density $\mathcal{J}(\omega)=\sum_{\bf k} |g_{\bf k}|^2\delta(\omega-\omega_{\bf k})$, and 
reorganization energy $E_R=\int_0^{\infty} d \omega \mathcal{J}(\omega)/{\omega}$, quantifying the form and strength of 
the system-bath coupling, respectively. Hence, the system-bath Hamiltonian reads $H_{SB} =\sum_{m=1}^{M}\sigma^{+}_{m} 
\sigma^{-}_{m}\otimes B_m({\bf r}_m)$, with bath operators written in the interaction picture as $B_m({\bf r}_m, 
t)=\sum_{\bf k}|g_{\bf k}|\left(b_{\bf k}^{\dagger} e^{i(\omega_{\bf k}t+{\bf k} \cdot {\bf r}_m)} +b_{\bf k} e^ 
{-i(\omega_{\bf k}t+{\bf k} \cdot {\bf r}_m)}\right)$.

Here we consider the limit of weak system-bath interaction for which the energy transfer dynamics can be described with 
a master equation derived within the Born-Markov and rotating-wave approximations \cite{breuerbook, masoud08}: 
$\dot{\rho}=-i [H_S+H_{LS}, \rho] + {\mathcal D}(\rho)$. Here, $\rho$ is the reduced system density operator while 
$H_{LS}$  is the Lamb-shift, which will be neglected.
The dissipator, ${\mathcal D}(\rho)=\sum_{\omega}\sum_{m,n}\gamma_{mn}(\omega, d_{mn}) {\mathcal F}_{mn}(\omega, \rho)$, 
accounts for the effects of both local and correlated fluctuations on the system. 
The operator ${\mathcal F}_{mn}(\omega, \rho) = A_n(\omega)\rho(t) 
A_m^\dag(\omega)-\frac{1}{2}\{A_m^{\dag}(\omega)A_n(\omega),\rho\}$, is written in terms of Lindblad operators 
$A_m(\omega)$, with $\{\omega\}$ the frequency spectrum given by the energy differences between single-excitation 
eigenstates of $H_S$. The 
rates $\gamma_{mn}(\omega, d_{mn})=\int_{-\infty}^{\infty}dt e^{i\omega t}C_{mn}(t, d_{mn})$, where 
$d_{mn}=|{\bf r}_m-{\bf r}_n|$ is the separation of sites $m$ and $n$, are defined in terms of bath correlation functions 
$C_{mn}(t, d_{mn})=\langle B_m({\bf r}_m, t)B_n({\bf r}_n, 0)\rangle_B$, with the average taken over the thermal 
equilibrium state of the bath. 
Evaluating the correlation functions, we obtain a formula relating the spatially-correlated dephasing rates to the 
on-site rates,
$\gamma_{mn}(\omega, d_{mn})={\mathcal G} (k(\omega)d_{mn})\gamma_{mm}(\omega)$, where the single-site rate 
is $\gamma_{mm}(\omega)=\gamma(\omega)=2\pi {\mathcal J} (|\omega|)|N(-\omega)|$, with $N(\omega)$ being the thermal 
occupation number.
The function $\mathcal{G}(z)$ depends on the environment dimension, exhibiting similar decaying behaviors in two and 
three dimensions \cite{doll06}.  Here, we assume a two-dimensional bath such that $\mathcal{G}(z)= J_0(z)$, with 
$J_0(z)$ being a Bessel function of the first kind. The function $k(\omega)$ defines an effective bath correlation 
length, which as a first approximation we assume to have a frequency-independent form $k(\omega)=1/R_B$, 
accounting for a phenomenological correlation length $R_B$ \cite{renger98}. We can now simplify the dissipator as 
${\mathcal D}(\rho)=\sum_{\omega}\gamma(\omega)\big[\sum_{m}{\mathcal F}_{mm}(\omega,\rho)
+\sum_{n\neq m} J_0(d_{mn}/R_B) {\mathcal F}_{mn}(\omega, \rho) \big]$, where the first term accounts for 
independent, on-site fluctuations, and the second for correlated processes. Using the property $\sum_{m,n}{\mathcal 
F}_{mn}(\omega, \rho)=0$, we can demonstrate that for $R_B \rightarrow \infty$, we have ${\mathcal 
D}(\rho)\rightarrow0$. 
In this limit all sites are fully correlated, and the single-excitation subspace of $H_S$ is effectively decoupled from 
the bath. In the opposite limit of vanishing $R_B$,  we recover the case that each site experiences only local 
fluctuations, ${\mathcal D}(\rho)=\sum_{\omega,m}\gamma(\omega){\mathcal F}_{mm}(\omega,\rho)$. 

Over the course of the system's evolution any initial excitation will eventually be either dissipated or trapped, with 
respective rates $\Gamma_m$ and $\kappa_m$ for each site.  These two incoherent processes can be accounted for by 
the inclusion of a non-hermitian operator, $H_{\rm proj}=-i\sum_m(\Gamma_m +\kappa_m)\sigma^{\dag}_{m}\sigma^{-}_{m}$, 
into the system Hamiltonian \cite{alexandra08, masoud08}. The efficiency of transfer (quantum yield) is then defined as 
the total probability that the excitation is trapped,
$\eta =  \sum_m \int_0^\infty \kappa_m \langle m |\rho(t)|{m}\rangle  dt$, where $|m\rangle=\sigma_m^+|0\rangle$.


\begin{figure}
\centering
\includegraphics*{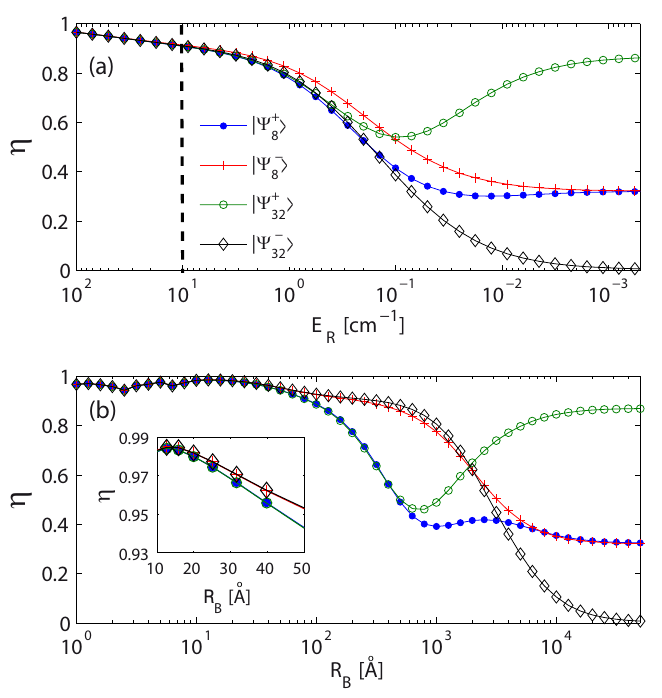}
\caption{Transfer efficiency at room temperature ($293$~K): (a) versus on-site reorganization energy 
$E_R$ in the absence of  correlated dephasing. $E_R$ increases from right to left along the horizontal axis to allow 
easier comparison with below; (b) versus $R_B$ for $E_R=100$~cm$^{-1}$.} 
\label{fig:erlb}
\end{figure}
{\it Results} - 
In the following, we assume that all sites dissipate at a rate $\Gamma=1~{\rm ns}^{-1}$ \cite{sener05}, with the 
exception of the special pair at the RC, which traps excitation at a rate $\kappa= 4~{\rm ps}^{-1}$ \cite{sener05}. For 
LH complexes the system-bath interaction is usually characterized by an Ohmic spectral density with either Drude 
\cite{rutkauskas06} or exponential cutoff \cite{renger98}; we assume a Drude 
form, $\mathcal{J}(\omega)=2E_R\omega_c \omega/\pi (\omega_c^2+\omega^2)$. Typical $E_R$ values are estimated to be 
between $100$~cm$^{-1}$\cite{renger02} and $200$~cm$^{-1}$\cite{freiberg09} for LH1, while the bath cutoff frequencies 
$\omega_c$ used to reproduce experimental data 
range from $50$~cm$^{-1}$\cite{renger02} to $1000$~cm$^{-1}$\cite{rutkauskas06}. Here, 
to be consistent with our weak system-bath coupling model and to unambiguously illustrate how initial state 
distinguishability can 
arise through spatial correlations, we consider a fast-modulated bath with $\omega_c=300\ {\rm cm}^{-1}$, and a maximum 
$E_R=100\ {\rm cm}^{-1}$. 
The ring sites are numbered from $1$ to $32$ (see \ref{fig:ring}) and we assume pure initial states, having no overlap 
with the RC, of the form $|\Psi_m^{\pm}\rangle=(1/\sqrt{m})\sum_{j=1}^m (\pm 1)^j |j\rangle$, where $1\leq m \leq 32$. 
The initial state is thus symmetrically or asymmetrically delocalized over the ring. In particular, we focus on four 
representative states $|\Psi_{8}^{+}\rangle$, $|\Psi_{8}^{-}\rangle$, $|\Psi_{32}^{+}\rangle$, and 
$|\Psi_{32}^{-}\rangle$, which in the case of pure coherent evolution ($R_B\rightarrow \infty$ or $E_R=0$) have, 
in general, different transfer efficiencies \cite{alexandra08}. Further, $|\Psi_{32}^{-}\rangle$ 
($|\Psi_{32}^{+}\rangle$) has a large degree of overlap with the lowest (highest) eigenstate of $H_S$. Since the 
dynamics of $\rho$ varies 
with respect to pigment-exchange, the results presented correspond to an average over all initial states with 
delocalization over $m$ consecutive sites.

In \ref{fig:erlb}(a), we plot the transfer efficiency, $\eta$, for each initial state as a function of the system-bath 
coupling quantified by $E_R$,  and accounting only for local dephasing, i.e. $\gamma_{mn}(\omega)=0$ for $m \neq n$. For 
small $E_R$, $\eta$ varies significantly between the different initial states, consistent with the coherent limit 
\cite{alexandra08}. As $E_R$ increases, exciton relaxation and dephasing take place and above $E_R\gtrsim10$~cm$^{-1}$ (dashed line), 
the efficiency becomes practically identical 
for all initial states considered, indicating that energy transfer is dominated by incoherent 
processes. Notice that $E_R\sim10$~cm$^{-1}$ is comparable to the electronic interaction between a Bchl in the ring and the special pair in the center \cite{alexandra08}.  
The situation differs for correlated dephasing, as shown in \ref{fig:erlb}(b). Here we fix 
$E_R=100$~cm$^{-1}$, which falls in the ranges of reported values for an LH1 \cite{renger02}, and plot $\eta$ as 
a function of the bath correlation length $R_B$.  For large enough values 
of $R_B$ the efficiency distinguishes again the initial state meaning that coherent 
dynamics is at play. Two key features can be drawn out from the comparison of 
\ref{fig:erlb} (a) and (b). First, for moderate values of $R_B$ the transfer efficiency clearly separates 
symmetric from asymmetric states - we shall shortly discuss the role of coherence and correlations in such 
distinguishability. Second, for an extended system, increasing $R_B$ does \emph {not} generally correspond simply 
to a renormalized coupling to an effective local environment at each site. This contrasts with the intuition to be 
gained by considering the two-site problem ($M=2$). 
In this case, we again use $\sum_{m,n}{\mathcal F}_{mn}(\omega,\rho)=0$ to rewrite the dissipator as 
${\mathcal D}(\rho)=\sum_{\omega}\gamma(\omega)[1- J_0(d/R_B)]\sum_{m=1,2}{\mathcal F}_{mm}(\omega, \rho)$, where 
$d$ is the inter-site separation. Here, a finite $R_B$ is equivalent to considering each site to be independently 
coupled to its own local bath, but with an effective reorganization energy ${\tilde E}_{R}=E_R[1- J_0(d/R_B)]$.

\begin{figure}
\centering
\includegraphics*{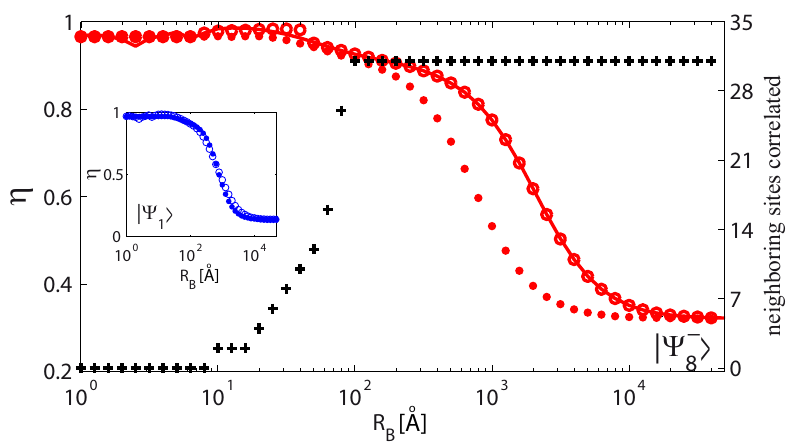}
\caption{Main: (Left axis) Room temperature transfer efficiency for $|\Psi_8^{-}\rangle$ against 
$R_B$, with $E_R=100$~cm$^{-1}$: ($\bullet$) assuming local dephasing with an effective $\tilde{E}_R=E_R[1-\langle 
J(d_{mn}/R_B)\rangle]$; (o) including only correlations satisfying $J(x)\geq0.7$; (line) including all 
correlations. (Right axis) ($+$) Effective number of correlated neighbors to each site. Inset: $\eta$ versus $R_B$ 
for the state $|\Psi_1\rangle$.} 
\label{fig:lbcomp}
\end{figure}
\begin{figure}
\centering
\includegraphics*{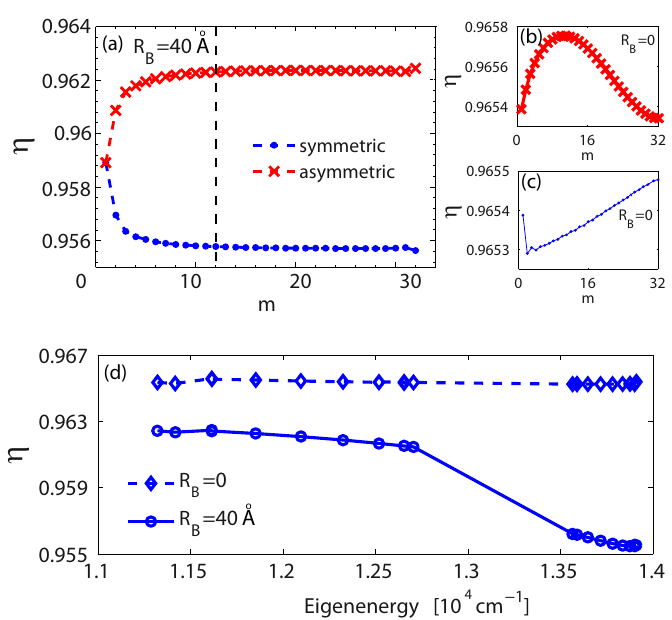}
\caption{(a) Efficiency for asymmetric and symmetric states as a function of delocalization length $m$ 
for $R_B=40$~$\mathring{A}$. (b-c) show $\eta$ in the absence of correlations ($R_B=0$) for asymmetric (b) and symmetric (c) states. (d) Efficiency for 
eigenstates of $H_S$ having vanishing overlap with the RC for $R_B=40$~$\mathring{A}$ ($\circ$), and $R_B=0$ 
($\diamond$). In all cases $E_R=100$~cm$^{-1}$.}
\label{fig:etavm}
\end{figure}
Interpretation of the effects of correlated dephasing therefore requires us to go beyond the two-site model. We analyze 
three different regimes: (i)  $R_B< d_{min}$, with $d_{min}\simeq 10$~$\mathring{A}$ the minimum pair distance in 
the system; (ii) $R_B\gg d_{max}=2R$, where $R\simeq 50$~$\mathring{A}$ is the radius of the ring; and (iii) the 
most interesting, intermediate regime $d_{min}< R_B <d_{max}$.  In the first two cases the two-site intuition can 
be applied as fluctuations are only very weakly correlated in case (i), while they are almost completely correlated at 
all sites in case (ii). In neither case is there significant variation from the average correlation over all $m\neq n$ 
pairs, $\langle J_0(d_{mn}/R_B)\rangle$. Hence, the dissipator has approximately the same form as the two-site 
case with an effective reorganization energy $E_R[1-\langle J_0(d_{mn}/R_B)\rangle ]$. This simplified dissipator 
does not reproduce the transfer efficiency in the intermediate regime for delocalized states, as can be seen in 
\ref{fig:lbcomp} where we plot $\eta$ for the initial state $|\Psi_8^-\rangle$ as an example.  Interestingly, for the 
state $|\Psi_1\rangle$, which has no quantum superpositions, the effective local-bath 
approximation captures the efficiency behavior across the whole range of $R_B$ (see \ref{fig:lbcomp} inset). The 
key feature, then, is that there is a crucial interplay between coherence, bath correlations, and initial state 
properties in defining the transfer efficiency. 
To understand this interplay we can follow an analysis within the quantum jump picture. The system dynamics can be seen 
as a series of coherent evolutions interrupted by incoherent jumps between single-excitation eigenstates, described by 
the action of the operator $\gamma_{mn}(\omega)A_m(\omega)\rho(t)A_n^{\dag}(\omega)$.  In the presence of bath 
correlations, the overlap with the trapping sites due to such incoherent jumps carries information on the quantum phase 
differences between sites $m$ and $n$, 
through a factor $(\pm 1)^{m+n}$, while in the absence of correlations it does not. Hence, after a jump due to a 
correlated bath, the subsequent coherent evolution is weighted by a phase factor that is always $1$ for symmetric 
states, but can be $1$ or $-1$ for asymmetric states. This gives rise to the symmetry distinguishability observed in 
\ref{fig:erlb}(b). In short, correlated fluctuations promote an {\it initial state symmetry dependent energy transfer} 
towards the RC. 

Further insight into the behavior of $\eta$ in the intermediate regime (iii) can be gained by associating $R_B$ to 
an effective number of neighbors correlated to each site. We estimate this number by neglecting all correlations below a 
cutoff $y$ ($J_0(d_{\alpha}/R_B)< y$), chosen to be the largest value that still gives a good agreement with the 
exact results. For the relevant points in \ref{fig:lbcomp}, we take $y=0.7$, which corresponds to neglecting all 
correlations between pairs whose distances satisfy $d_{mn}>1.1R_B$.  In Ref. \cite{rutkauskas06} correlations 
among $5$ consecutive sites were included to fit measurements of optical properties of LH1 samples. From the above 
analysis, we can associate this number to a correlation length of $R_B \simeq 20$~$\mathring{A}$ (see 
\ref{fig:lbcomp}).  Notice that at this moderate correlation length, $\eta$ already starts distinguishing states 
according to their symmetry (inset in \ref{fig:erlb}(b)). 

With $R_B$ now associated to an effective number of correlated neighbors we are led to conjecture that, for a 
given $R_B$, the interplay with coherence will be such that $\eta$ should increase as a function of the 
delocalization length ($m$) for asymmetric states, while it should decrease for symmetric states. Further, such behavior 
should happen only up to an {\it optimal delocalization length} $m_{c}$, beyond which the efficiency should not 
considerably change. This is indeed the behavior seen in \ref{fig:etavm}(a), where we plot $\eta$ for asymmetric and 
symmetric initial states as a function of $m$, for an intermediate value $R_B=40$~$\mathring{A}$ with associated 
$m_c\simeq 12$ (dashed line). These results confirm that the interplay of coherence and correlations allows the system 
to exploit not just asymmetry, but also exciton delocalization. We find, however, that the optimal $m_c$ for which the 
efficiency plateaus increases only slightly with larger $R_B$, and is never greater than $m_c\simeq 16$.  Hence, 
$m_{c}$ is not simply dictated by the length of correlations, but is mainly defined by the symmetry of the electronic 
interactions in the LH1-RC core. Further, the marked splitting between the two sets of states around the efficiency 
associated to $m=1$ clearly illustrates the symmetry-dependent energy transfer discussed above. To highlight the 
difference to the uncorrelated environments case, \ref{fig:etavm}(b-c) show $\eta$ as a function of $m$ when 
$R_B=0$.  Importantly, although the variation in $\eta$ is negligible, its functional form for each set of states 
is the same as that observed under purely coherent evolution \cite{alexandra08}.   

In the actual light-harvesting process the implications of the above discussed phenomena may be rather subtle. To 
illustrate this, in \ref{fig:etavm}(d) we plot $\eta$ against energy for all eigenstates having vanishing overlap with 
the RC. The presence of correlations is manifested in an unexpected ordering of the eigenstates whereby $\eta$ decreases 
monotonically with increasing energy. Such order is not observed when correlations vanish.  In a functioning 
photosynthetic system the LH1 donor state will be a thermal distribution about the lowest energy levels \cite{hu97}. Our 
results indicate that, without correlations, such a distribution will be as efficient as one about higher states. In 
contrast, if correlations are present, a thermal distribution about low-lying states would be the best option. This 
suggests that the interplay of coherence and correlations could indeed help the natural system to differentiate the best 
energy transfer pathway.

In summary, we have shown how coherence and bath correlations could equip light-harvesting systems with a mechanism to 
exploit  initial state phase information at room temperature. Although our conclusions are based on a simple model, we conjecture that 
initial state distinguishability in energy transfer may be a general feature when spatial and/or temporal environmental 
correlations are present. Hence, experimental explorations of such phenomena could help in elucidating 
both microscopic details of the environmental correlations as well as the advantages of coherent dynamics in natural and artificial light-harvesting systems~\cite{collini10,womick09}.

\acknowledgement
We thank T. Osborne and D. Porras for interesting discussions.  F. F. thanks CONICYT  for 
support. A.N and A.O-C acknowledge funding from the EPSRC.

\subsection{Supporting Information Available}
Transfer efficiency at room temperature as a function of  $R_B$ including static disorder. It is shown that for
$R_B<2R$, moderate static disorder has no significant effect on the results presented.\\
This material is available free of charge via the Internet at http://pubs.acs.org.

\end{document}